\documentclass[a4paper]{JHEP3}

\JHEPspecialurl{}

\usepackage{epsfig,multicol,bbm}

\newcommand\fverb{\setbox\fverbbox=\hbox\bgroup\verb}
\newcommand\fverbdo{\egroup\medskip\noindent%
            \fbox{\unhbox\fverbbox}\ }
\newcommand\fverbit{\egroup\item[\fbox{\unhbox\fverbbox}]}
\newbox\fverbbox

\title{Black hole complementarity with local horizons and Horowitz-Maldacena's proposal}

\author{Sungwook E. Hong, Dong-il Hwang, Dong-han Yeom and Heeseung Zoe\\
    Department of Physics, KAIST,\\ Daejeon, 305-701, South Korea\\
    E-mail: \email{eostm@muon.kaist.ac.kr}, \email{eastone83@gmail.com}, \email{innocent@muon.kaist.ac.kr}, \email{hszoe@muon.kaist.ac.kr}}

\received{\today}
\revised{}
\accepted{}

\abstract{
To implement the consistent black hole complementarity principle,
we need two assumptions: first, there exists a singularity near the center, and second, global horizons are the same as local horizons.
However, these assumptions are not true in general.
In this paper, the authors study a charged black hole in which the second assumption may not hold.
From the previous simulations, we have argued that the event horizon is quite close to the outer horizon, and it seems not harmful to black hole complementarity;
however, the Cauchy horizon can be different from the inner horizon, and a violation of complementarity will be possible.
To maintain complementarity, we need to assume a selection principle between the singularity and the Hawking radiation generating surface;
we suggest that Horowitz-Maldacena's proposal can be useful for this purpose.
Finally, we discussed some conditions under which the selection principle may not work.
}

\keywords{Black Holes, Black Holes in String Theory}

\dedicated{}

\begin{document}

\section{\label{sec:intro}Introduction}

The black hole information loss problem is a very important issue to consider when trying to understand the principles of quantum gravity \cite{inforpara}.
There have been many proposals to solve the problem \cite{Susskind:2002ri}, and one of the most important proposals is the black hole complementarity principle \cite{complementarity}
which is supported by the membrane paradigm \cite{Thorne:1986iy} and the holographic principle \cite{Susskind:1994vu}.

The main argument of black hole complementarity is as follows \cite{complementarity}:
an asymptotic observer, who is on the outside of the black hole, will observe all information from the Hawking radiation,
and there is no violation of natural laws, e.g., the unitarity of quantum mechanics.
But a free-falling observer, who is going to pass through the black hole horizon, will observe that all information falls into the singularity;
here also, there is no violation of natural laws.
However, one may guess that the information is duplicated, since one set of information is observed from the outside, and the other set of information is observed from the inside of the black hole.
That is impossible because of the no cloning theorem in quantum mechanics.
Now, black hole complementarity argues that if the two observers cannot communicate forever, there is essentially no problem.
This resolves the black hole information loss problem in a fascinating way.

The black hole complementarity principle is related to the holographic principle \cite{Susskind:1994vu}.
Moreover, according to AdS/CFT \cite{Maldacena:1997re}, if we assume the background AdS,
all processes that happen inside of the bulk anti de Sitter (AdS) space can be described by the conformal field theory (CFT) of boundary.
Since the CFT is manifestly unitary, we can think that black hole physics must be unitary and information must be conserved.
We know that, \textit{if we assume the entropy formula} ($S=A/4$) \cite{Strominger:1996sh} \textit{and the unitarity} \cite{Maldacena:1997re}, the information must escape around the information retention time \cite{Page:1993df};
and, at this time, the black hole can be still large enough and semi-classical.
Thus, somehow the black hole complementarity principle plus the nonlocal effects seem to be manifest in semi-classical black holes \cite{nonlocal}.

However, what if one observer observes the information from the Hawking radiation, and finally falls freely into the black hole?
Then, he or she may compare two copies of the information, and conclude that the information was duplicated.
Then one may think that black hole complementarity must have a fatal problem.
To prevent this situation, black hole complementarity insists on two key assumptions:
one is that \textit{the inner structure of black hole contains a singularity};
the other is that \textit{the event horizon is functioning as a local horizon that generates Hawking radiation}.

If we assume those two things, we can resolve the duplication paradox.
If the initial information falls beyond the event horizon, because of the causal definition of the event horizon,
the duplication cannot be observed on the \textit{outside} of the black hole.
Moreover, since the in-falling information will touch the singularity before the information retention time \cite{Page:1993df},
the duplication is hardly observed on the \textit{inside} of the black hole;
the in-falling information needs an energy order of $\exp{M^{2}} \gg M$ to send the signal to the observer for a black hole of mass $M$,
but this seems to be impossible \cite{inforretention}\cite{inforretention2}.
These are the main speculations of the inventors of the black hole complementarity principle.

However, those two assumptions are not valid in generic types of black holes.
For a regular black hole that has no singularity inside,
two of authors have already shown that we can do a duplication experiment and that complementarity may not hold \cite{YZ}.

For a charged black hole that has two horizons, the problem becomes more difficult;
there was a naive speculation that it seems to be possible to observe the duplication beyond the inner horizon \cite{Ge:2005bn}.
However, if the inner horizon is the same as the Cauchy horizon, since Cauchy horizon means the limitation of our knowledge,
one may think that it is not harmful to black hole complementarity.
Notice that, in this case, someone also used a device similar to the second assumption; \textit{the inner horizon is the same as the Cauchy horizon.}

In this paper, we discuss how black hole complementarity should work in a dynamical charged black hole.
We start the speculation from a proper interpretation of the Penrose diagram for such a black hole, and confirm it with previous simulation results \cite{HHSY}.
However, we will not go into the details of the simulation (see Appendix \ref{appa}).
It should be remarked that we need to assume a large number of massless fields (or large $N$ limit, where $N$ is the number of massless fields) to resolve the trans-Planckian curvature problem. \footnote{Quantitatively, $N$ is needed on the order of $\exp{100}$ for our simulations. For quantitative comments on the number, as well as on how to resolve the curvature problem by using large $N$, see Appendix \ref{appa}.}

In Section \ref{sec:complementarity}, we introduce two requisites for black hole complementarity. And, we discuss how, if we drop one of these, the observation of information duplication seems to be possible.
In Section \ref{sec:general}, we discuss the causal structure of a dynamical charged black hole.
In Section \ref{sec:global_local}, we discuss two possible scenarios that may be harmful to black hole complementarity.
First, we discuss the case in which the event horizon is different from the outer horizon;
and from previous simulations, this seems not harmful to black hole complementarity.
Second, we observe that the Cauchy horizon can be different from the inner horizon, and that this situation will be harmful to black hole complementarity.
In Section \ref{sec:HM}, we discuss that the Horowitz-Maldacena proposal can be helpful to rescue black hole complementarity,
since it can be a selection mechanism between the singularity and the Hawking radiation generating surface.
Finally, in Section \ref{sec:dis}, we summarize our discussion.

\section{\label{sec:complementarity}Two requisites for black hole complementarity}

Among many ideas suggested to resolve the black hole information paradox,
black hole complementarity was motivated from the success of string theory in reproducing the entropy formula \cite{Strominger:1996sh}\cite{Callan:1996dv}\cite{Maldacena:1996ky}
and in developing the holographic principle \cite{Susskind:1994vu}\cite{Maldacena:1997re}\cite{inforretention2}.
Now, if one wants to say that the complementary argument should be a generic principle of black hole physics, we should clarify what we need.
Thus, as we discussed in the introduction, we need two assumptions to maintain the complementarity:
\begin{itemize}
  \item The inner structure of the black hole contains a singularity;
  \item The global horizons (event and Cauchy horizons) are the same as the local horizons (outer and inner horizons); thus, the event horizon is functioning as a Hawking radiation source, and the inner horizon is functioning as the boundary of the predictable region.
\end{itemize}

Note that, as we commented in the introduction, if we assume these two facts, there is no trouble with black hole complementarity.

However, in general, those two assumptions are not true.
For the first assumption, it is possible to construct a regular black hole, i.e., a black hole without a singularity,
and the authors have already discussed this possibility with the duplication problem \cite{YZ}.
We will not discuss this assumption further in this paper.

It is very interesting that the second assumption is not true generally, either.
The event horizon is defined in a causal sense \cite{Hawking:1973uf}.
We do not know where the event horizon is just before we know the whole causal structure from past infinity to future infinity.
From this problem, some authors have developed useful concepts of local horizons to define the black hole in terms of local geometry:
the apparent horizon, the dynamical horizon, the isolated horizon, or the trapping horizon \cite{Ashtekar:2004cn}.

In static cases, the event horizon is the same as the locally defined horizons.
However, in dynamic cases, they cannot be same.
Those definitions have some merits and demerits case by case,
but in our charged black hole case, the apparent horizon will be useful to define the outer and the inner local horizons (See \cite{Hawking:1973uf}; for the spherically symmetrical case, it is defined by $r_{v} (\equiv \partial r/ \partial v ) = 0$, i.e., the radial function does not increase or decrease along the outgoing null direction).
There are concrete arguments that the globally defined event horizon is not related to the Hawking radiation,
but that the locally defined outer horizon is working as the Hawking radiation generating surface \cite{Visser:2001kq}\cite{Nielsen:2008kd}.

The event horizon may be different from the outer horizon.
\textit{If one can send a signal between the event horizon and the outer horizon, one may see the duplication from the outside of the black hole.}
This means that this may be harmful to black hole complementarity.

For the inner horizon, it was already pointed out that black hole complementarity could be invalid in a charged black hole \cite{Ge:2005bn}\cite{Thorlacius} or in some black holes that have two horizons \cite{Ge}.
However, in fact, the counterargument from the inner horizon is less clear, since it just studies the black hole by using only classical metric solutions; thus, it could not include the mass inflation effect \cite{Poisson:1990eh}, which is a characteristic phenomena of the inner horizon.
If one includes the mass inflation effect, it will make the inner horizon unstable, and the inner horizon may become a singularity \cite{Bonanno:1994ma}.
Then, black hole complementarity will be safe by the inner horizon.

But, it is still unclear whether the inner horizon must behave as a true (strong) singularity or not \cite{Ori};
so, if we can resolve the problem of the inner horizon by some (maybe not yet known) assumptions, one may see the duplication beyond the inner horizon.
Even if we can see the duplication beyond the inner horizon, however, \textit{if the inner horizon is the Cauchy horizon} from the second assumption,
from the definition of the Cauchy horizon \cite{Hawking:1973uf}, we cannot give any definite arguments about it;
for example, if the time-like singularity generates a strong thunderbolt \cite{Hawking:1992ti}, the semi-classical analysis should be breakdown, and we cannot say any definite things about the duplication experiment beyond the Cauchy horizon.
In this sense, even though there is no singularity along the inner horizon, as long as the inner horizon is the Cauchy horizon, the duplication experiment will be difficult in a charged black hole.

However, if we can generate a situation in which the Cauchy horizon is separated from the inner horizon,
we may see the duplication between the inner horizon and the Cauchy horizon; then, it can obviously be harmful to black hole complementarity.

To check those possibilities, we will study the general causal structure of a dynamical charged black hole.
Our work is crucially related to our previous simulations \cite{HHSY}, and we assumed large $N$.
We will briefly introduce previous results in the next section; and will discuss the complementarity problem.

\section{\label{sec:general}General causal structure of a dynamical charged black hole}

\begin{figure}
\begin{center}
\includegraphics[scale=0.55]{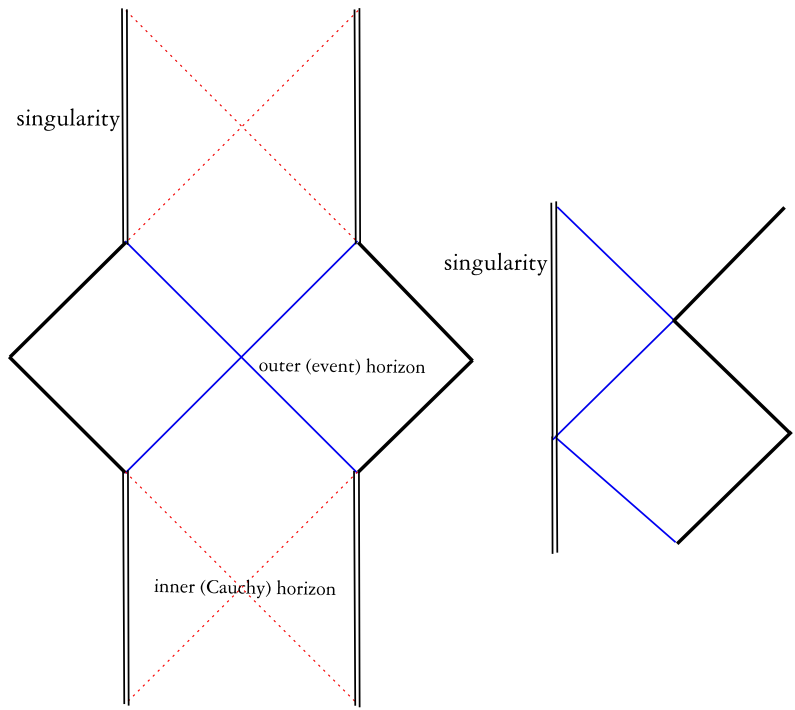}
\caption{\label{fig:charged_metric}The left figure is the Penrose diagram of a static charged black hole, for $M > Q$. The right diagram is the Penrose diagram of a static charged black hole, for $M = Q$. In this case, the outer horizon is the same as the inner horizon.}
\end{center}
\end{figure}

For a static case, the charged black hole solution is well-known \cite{Hawking:1973uf}. The following is the solution
\begin{eqnarray} \label{charged_metric}
ds^{2} = -\left(1-\frac{2M}{r}+\frac{Q^{2}}{r^{2}}\right) dt^{2} +\left(1-\frac{2M}{r}+\frac{Q^{2}}{r^{2}}\right)^{-1}dr^{2}+r^{2}d\Omega^{2},
\end{eqnarray}
where $M$ is the mass, and $Q$ is the electric charge.
From this metric, we can draw a maximally extended causal structure,
and we get the Penrose diagram for this static charged black hole solution (Figure \ref{fig:charged_metric}; \cite{Hawking:1973uf}).
However, these diagrams are different in dynamic situations.
First, the real situation does not have time symmetrical metric structures, and thus is similar to the Schwarzschild case.
The initial state is generally flat, and the final state will again be flat in many cases.

\begin{figure}
\begin{center}
\includegraphics[scale=0.6]{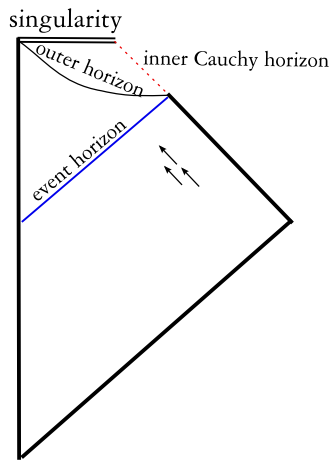}
\caption{\label{fig:charged_with_mass_inflation}Charged black hole with mass inflation. The collapsing matter generates the mass inflation and the space-like singularity.}
\end{center}
\end{figure}

Then, one may ask the question: what will happen if we push some charged matter into flat space-time?
One simple guess at a result is the growing of outer and inner horizons, and the evolving of a time-like singularity \cite{Vaidya}.
However, because of the mass inflation \cite{Poisson:1990eh}, this will not happen.
The inner horizon must have grown from a zero radius, but, because of the mass inflation, it cannot grow, and it becomes a central singularity; as a result, the central singularity becomes space-like \cite{Bonanno:1994ma}.
Thus, we may make a guess as seen in Figure \ref{fig:charged_with_mass_inflation}, and this has been confirmed by numerical simulations \cite{HodPiran}.

For more realistic situations, we need to include two effects: the pair creation effect and the Hawking radiation.
From some of the previous simulations, we have learned that, as long as the initial charged matter collapses quickly,
the discharging effect is no more dominant than the other effects during collapse \cite{SorkinPiran};
and, if we assume a large amount of charge, the pair-creation effect during evaporation will be exponentially suppressed \cite{inforretention2} (and, see also \cite{Ori:2001xc}; suppression of pair-creation can be obtained by choosing of a proper coupling).
This can be checked easily; the typical pair-creation rate $\Gamma$ is proportional to
\begin{eqnarray}
\Gamma \sim e^{-\pi\frac{m_{e}^{2}}{e E}} \sim e^{-\pi \frac{m_{e}^{2} M^{2}}{e Q}},
\end{eqnarray}
where $m_{e}$ is the electron mass, $e$ is the unit charge, and $E = Q/R^{2}$ is the electric field for a black hole of charge $Q$ and radius $R$ \cite{Dunne:2004nc}. Then, $E$ is approximately $Q/M^{2}$ for $M \gg Q$, and the above formula is obtained.
If $M \gg Q \gg e/m_{e}^2$, then
\begin{eqnarray}
\frac{m_{e}^{2} M^{2}}{e Q} \gg 1
\end{eqnarray}
holds, and approximately the pair-creation rate is proportional to $\exp{(-M)}$.\footnote{Therefore, in fact, if the mass of a black hole is sufficiently large, whether the charge is large or not, we can say that the pair-creation is small. And, if the charge is sufficiently large, we can say that, even if the black hole is near extreme limit, the pair-creation becomes exponentially suppressed.} One may notice that, for $M \gg Q$ cases, the change of mass for a unit time is proportional to $1/M^{2}$ from the Stephen-Boltzmann law, whereas the charge emission is suppressed exponentially. Then, our suppression assumption can be justified. Although $m_{e}^{2}/e$ is a small value in our universe, if large $Q$ and large $M \gg Q$ are allowed, our assumption can be justified \cite{inforretention2}; or, if we assume a universe in which all charged particles have a small charge, i.e., $m_{e}^{2}/e \sim 1$ holds, the pair-creation can be suppressed well.
Thus, by assuming some physical contents, we can derive a model in which the Hawking radiation drives the black hole to the extreme limit.
(Of course, we can think of a model in which the pair-creation is dominant; but, to discuss black hole complementarity, the pair-creation dominant model is not interesting.) Of course, we can also assume that this suppression can be maintained until the order of the information retention time.

Then, the Hawking temperature will be decreased, since the Hawking temperature of the charged black hole is \cite{Birrell:1982ix}
\begin{eqnarray} \label{Hawking_temperature}
T_{H} = \frac{1}{2 \pi} \frac{\sqrt{M^{2}-Q^{2}}}{(M+\sqrt{M^{2}-Q^{2}})^2}.
\end{eqnarray}
As the Hawking temperature decreases, the pair-creation rate and the mass loss rate approach the same value; then, the pair creation effect becomes important again \cite{SorkinPiran2}.
Then, maintaining $Q/M$ to be a constant value, the mass and charge will be decreased; and finally, the black hole will be changed from a charged one into a neutral one.
Then, because of the Hawking radiation, the black hole will totally evaporate, and we get flat space-time again.

\begin{figure}
\begin{center}
\includegraphics[scale=0.6]{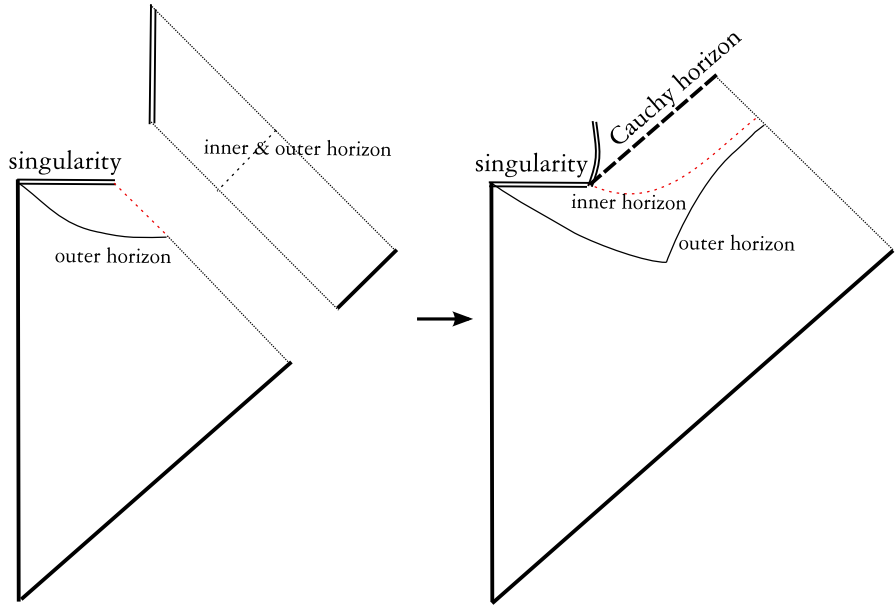}
\caption{\label{fig:charged_with_mass_inflation2}This represents the Penrose diagram to connect the mass inflation scenario to the extreme black hole solution.}
\end{center}
\end{figure}

Now, we need to draw smoothly for the generation and evolution of the inner horizon.
We need to connect the mass inflation scenario (Figure \ref{fig:charged_with_mass_inflation}) and the near extreme black hole (the right diagram in Figure \ref{fig:charged_metric});
the only possible way is that the inner horizon bends from constant $v$-direction to constant $u$-direction (Figure \ref{fig:charged_with_mass_inflation2}; \cite{SorkinPiran2}\cite{HHSY}).

\begin{figure}
\begin{center}
\includegraphics[scale=0.6]{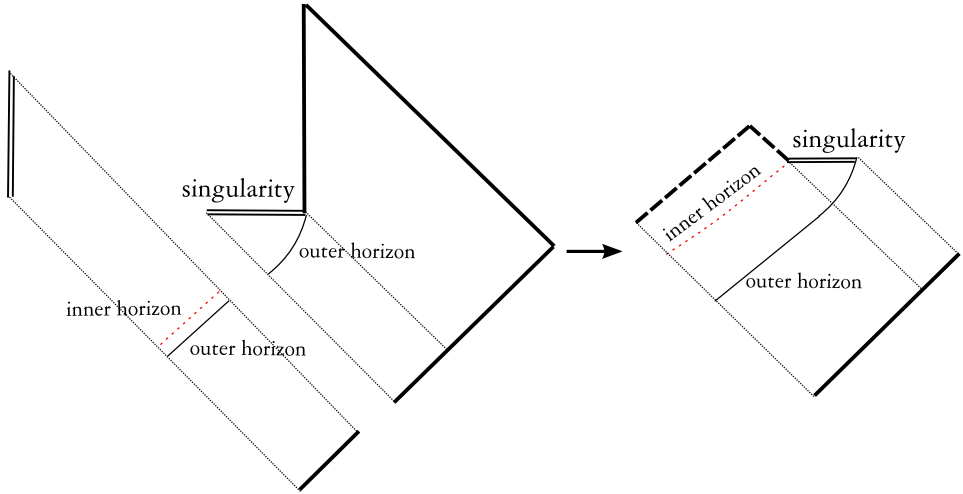}
\caption{\label{fig:charged_neutral}This represents the Penrose diagram to connect a near extreme black hole to a neutral black hole. In numerical calculation \cite{HHSY}, we pushed oppositely charged matter.}
\end{center}
\end{figure}

Also, we need to consider the charged-neutral transition; we need to connect the near extreme black hole and the neutral black hole (Figure \ref{fig:charged_neutral}). This diagram is less clear, since there will be two ways to connect the inner horizon and the space-like singularity. One possible way is that the inner horizon bends in a time-like direction; the other way is that the inner horizon remains in a space-like direction (analytic calculations on this issue are discussed in \cite{Levin:1996qt}). And, we observed that the latter scenario is one of the possible cases \cite{HHSY}. (This does not imply that the former scenario is impossible. It depends on the initial condition, and these analysis will be included in \cite{HHY}. In any case, this choice itself is not important to the discussion of black hole complementarity.)

\begin{figure}
\begin{center}
\includegraphics[scale=0.45]{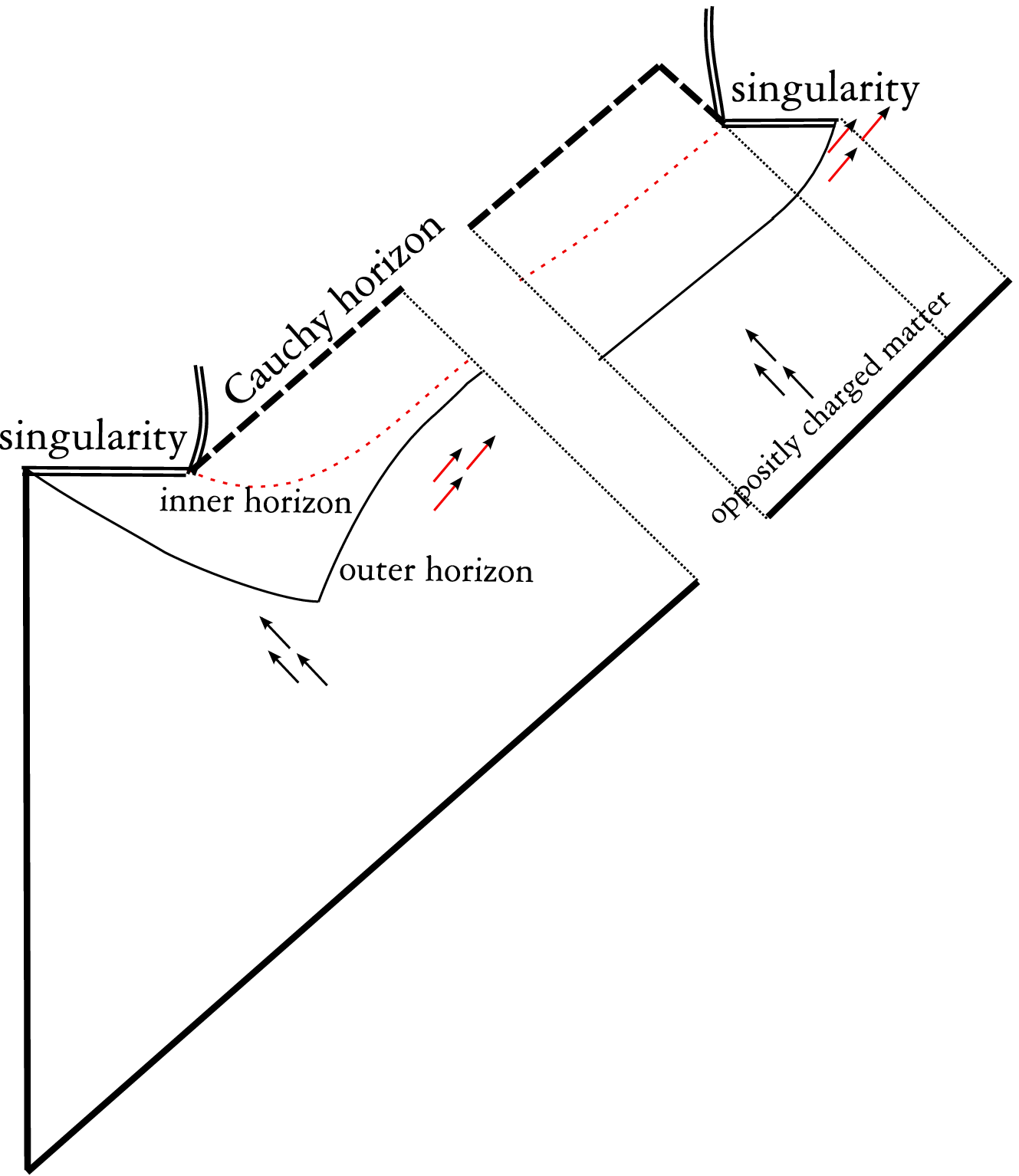}
\includegraphics[scale=0.55]{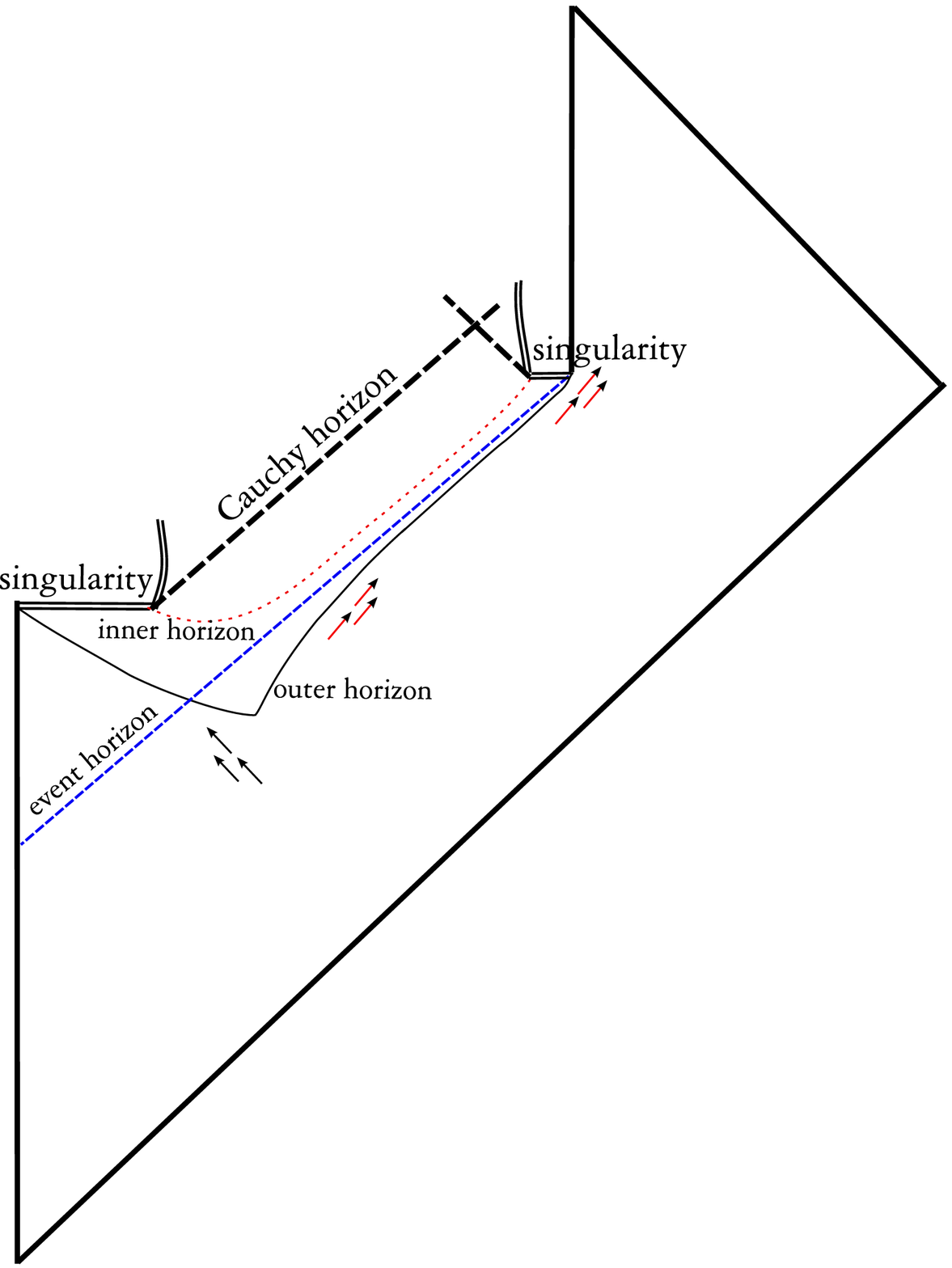}
\caption{\label{fig:charged} The lower part represents the Penrose diagram to connect the mass inflation scenario to the near extreme limit. The upper part represents the Penrose diagram for charged-neutral transition. If we connect the two diagrams, we get the causal structure of dynamical charged black holes.}
\end{center}
\end{figure}

Finally, we can draw a causal structure diagram by including the Hawking radiation and the pair-creation (Figure \ref{fig:charged}).
This result is consistent with some theoretical expectations on local horizons \cite{Ashtekar:2004cn} and is confirmed by previous simulations \cite{HHSY}.

One may ask whether the inner horizon is singular, as in Figure \ref{fig:charged_with_mass_inflation}.
However, from a simple calculation, we can check that the inner horizon should be regular,
since the mass inflation factor is proportional to $\exp(\kappa_{i} (u+v))$ \cite{Poisson:1990eh}
where $\kappa_{i}$ is the surface gravity of the inner horizon.
The inner horizon has a finite retarded time $u$ and a finite advanced time $v$;
thus, the mass inflation is finite for all finite $u$ and $v$.
It may become greater than the Planck scale,
but this problem can be regularized by assuming a large $N$.
If we assume a large $N$, it will re-scale the Planck length of our simulation, and it will change the Planckian cutoff in our simulation (see Appendix \ref{appa}).
In this paper, we will assume this. Then, the entire integrated domain (except near singularity) becomes semi-classical.

Note that, one can easily observe that the outer horizon is different from the event horizon,
as well as the fact that the inner horizon and the Cauchy horizon are separated;
thus, local horizons are different from global horizons.

\section{\label{sec:global_local}Global horizon vs. local horizon}

\begin{figure}
\begin{center}
\includegraphics[scale=0.85]{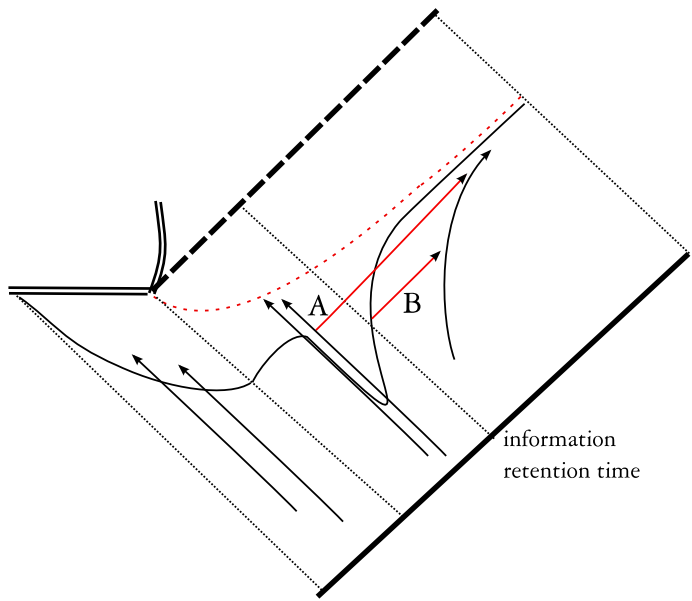}
\caption{\label{fig:complementarity_event_false} A potential but impossible duplication experiment. We send more matter to pull the outer horizon. The outer horizon will behave along the time-like direction; thus one may compare $A$ and $B$ outside of the black hole. However, it is improbable. See the next diagram.}
\end{center}
\end{figure}

\begin{figure}
\begin{center}

\includegraphics[angle=270, scale=0.4]{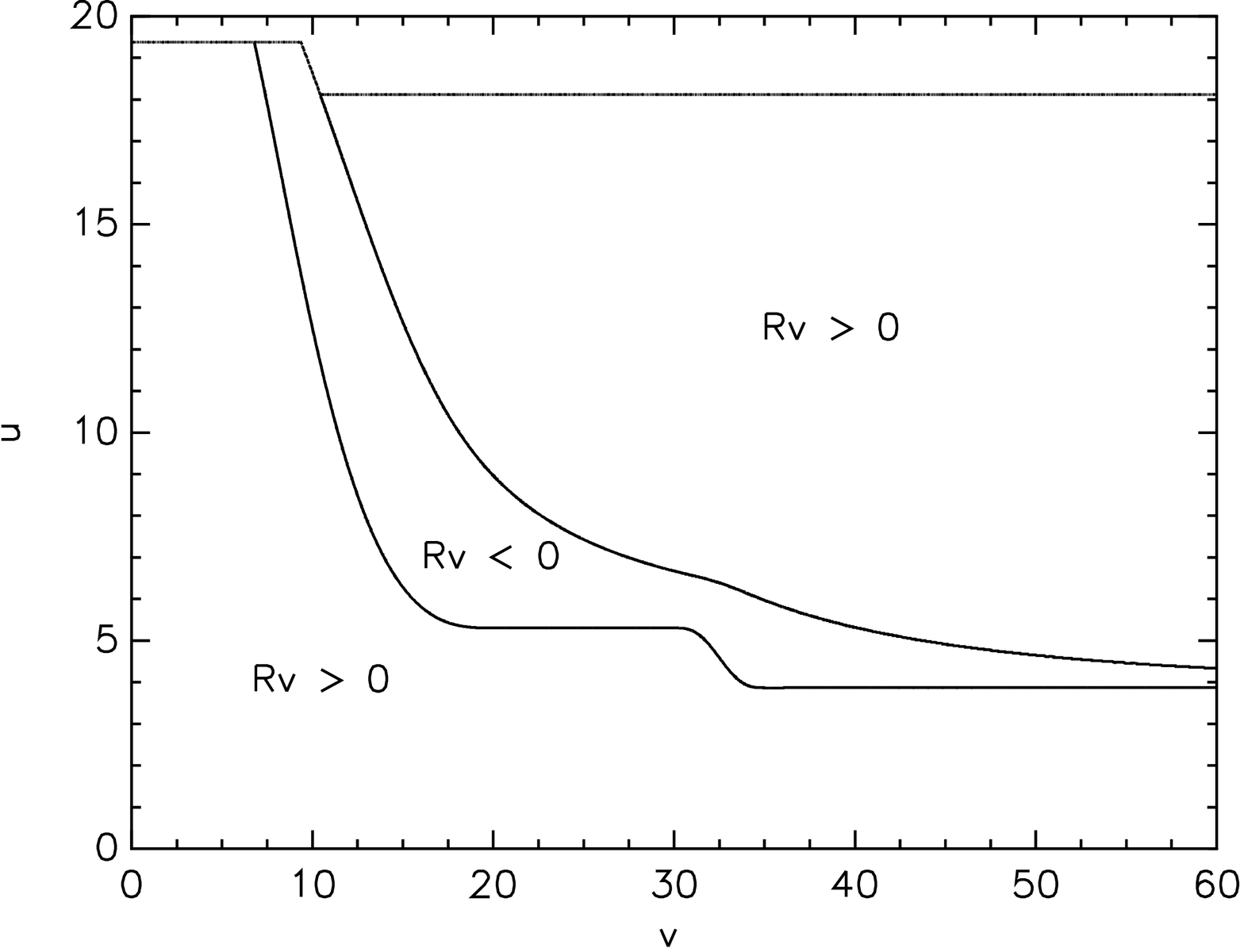}

\includegraphics[scale=0.85]{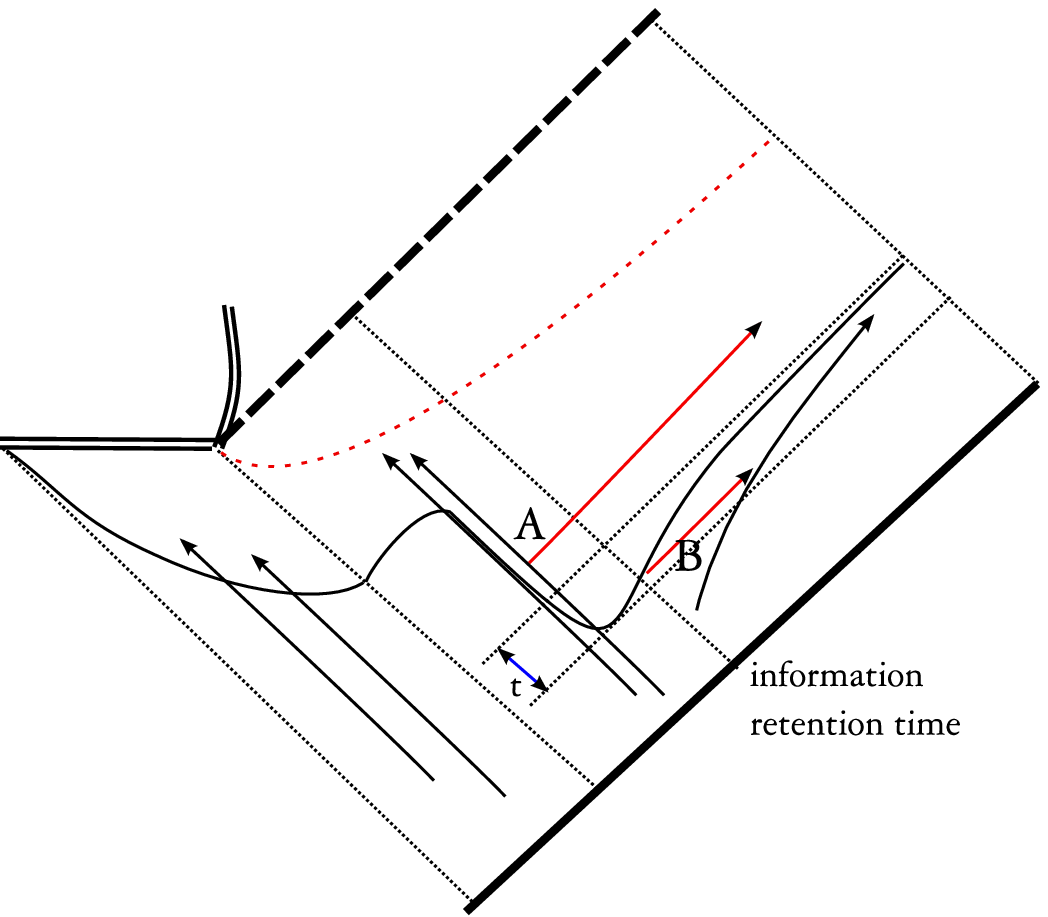}

\caption{\label{fig:complementarity_event_true} A simulation in which a strong pulse is inserted around $v=30$; this represents the apparent horizon $r_{v}=0$. The outer horizon behaves along almost the null direction; thus, the duplication seems to be improbable. This situation is quite similar to the Schwarzschild case. Compare with the next figure.}
\end{center}
\end{figure}

In this section, we will drop the second assumption of black hole complementarity.
Especially, we discuss two possible situations that may be harmful to the complementarity principle.
One possibility is to think that the event horizon is different from the outer horizon;
the other possibility is that the Cauchy horizon is different from the inner horizon.

\subsection{\label{sec:event_outer}The event horizon and the outer horizon}

Let's imagine a situation in which the outer horizon can be different from the event horizon.
Then, one may guess that a free-falling observer sends a signal to the out-going direction
before touching the event horizon; after the information retention time \cite{Page:1993df},
it can be observed by the outside observer, and the duplication may be observed (Figure \ref{fig:complementarity_event_false}).
One crucial problem is whether the distance between the event horizon and the outer horizon is large enough or not.
However, according to our simulation, this cannot happen;
the difference between the outer horizon and the event horizon is not sufficiently large, i.e., the situation is similar to the Schwarzschild case.

\begin{figure}
\begin{center}
\includegraphics[angle=270, scale=0.4]{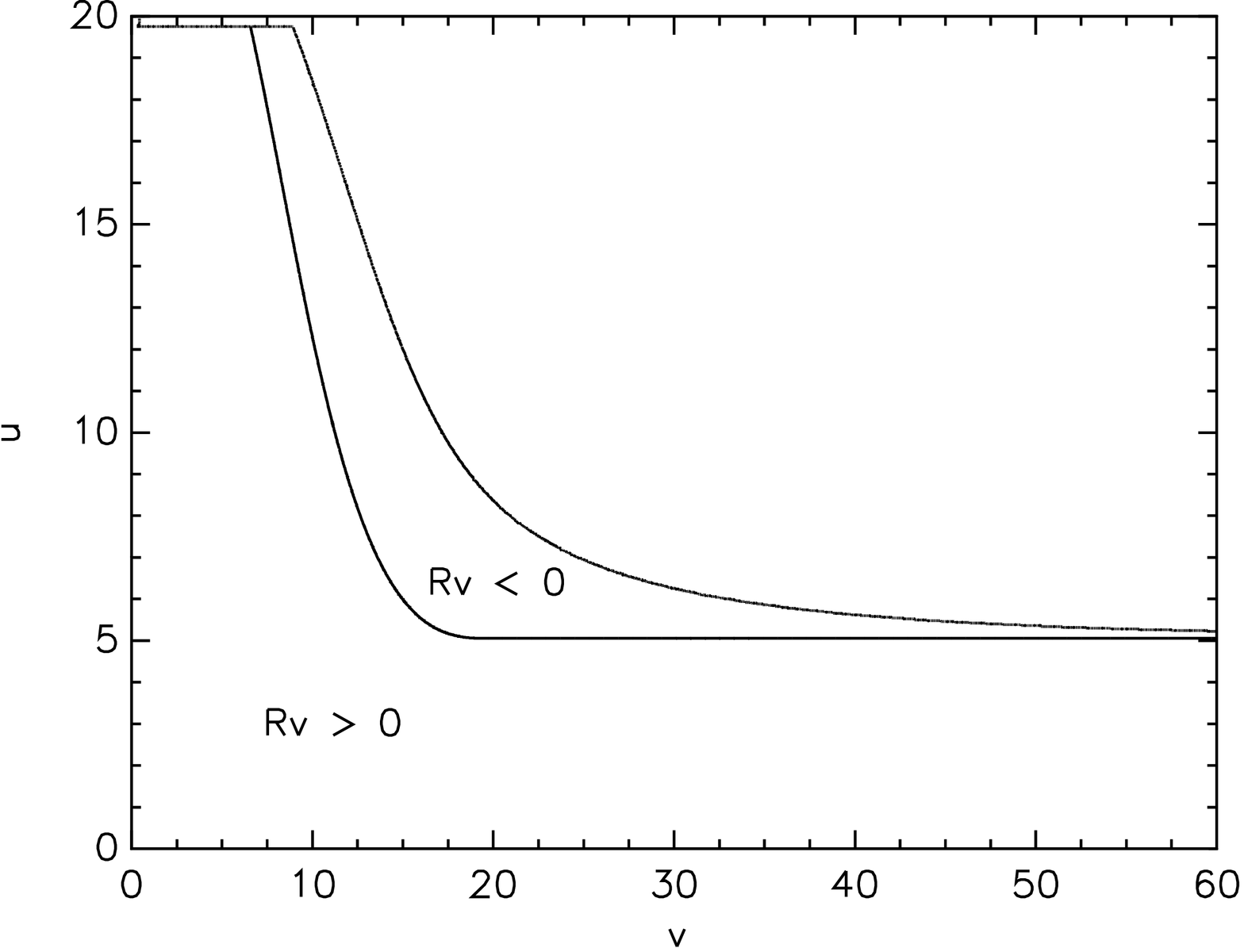}

\includegraphics[scale=0.8]{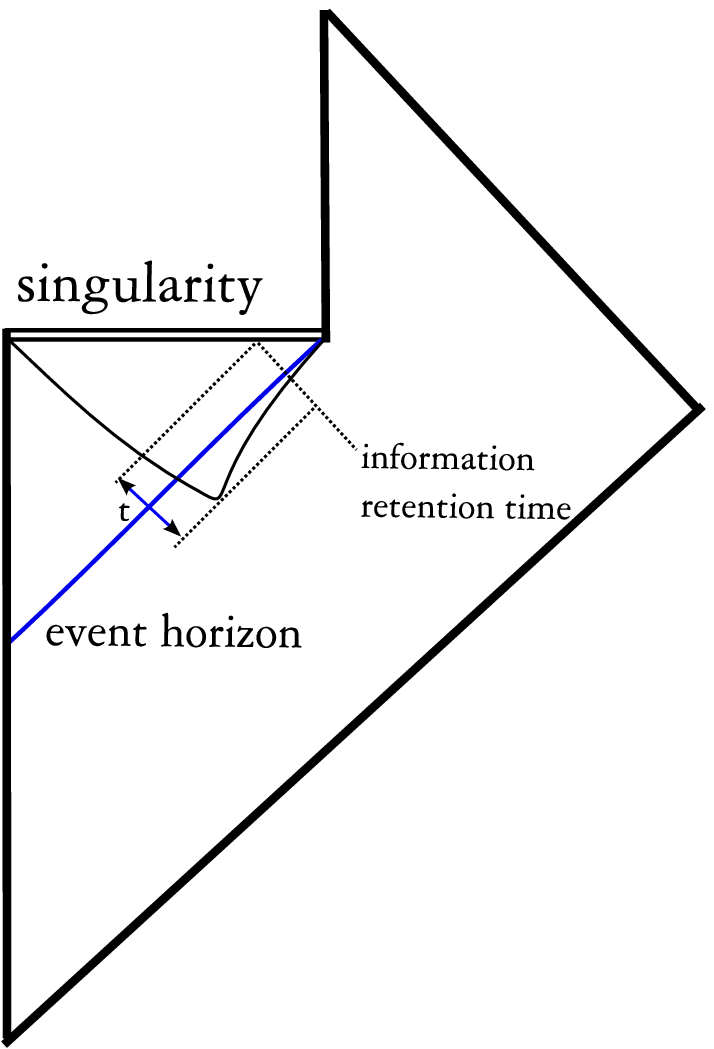}
\caption{\label{fig:Schwarzschild}This shows that, in a Schwarzschild case, the apparent horizon quickly approaches the null event horizon. In this case, $t$ could be calculated, and it was on the order of $\exp(-M^{2})$ with a black hole of mass $M$. Thus, the duplication experiment in this case is impossible; and, obviously, distance between the event horizon and the outer horizon is also too narrow. If we apply this argument to the case of a charged black hole, it seems to be improbable, too.}
\end{center}
\end{figure}

Let's compare Figure \ref{fig:complementarity_event_true} with a Schwarzschild case. Figure \ref{fig:Schwarzschild} shows that the outer horizon approaches the null event horizon. In both cases, they are very close, i.e., each $t$ are short.
For the latter case, it is known that it seems very difficult to do the duplication experiment even on the inside of the black hole, since the time scale is of an order of $\exp(-M^{2})$ \cite{inforretention}\cite{inforretention2}; therefore, of course, the distance between the event horizon and the outer horizon is more narrow, and it is improbably difficult to send information between them.
Then, it is reasonable to think that we can apply almost the same argument to the former case; thus, it may not harmful to the complementarity.
Or, it is fair to say that \textit{if one wants to find a counterexample of black hole complementarity, looking at the region between the event horizon and the outer horizon is not a good strategy.}

\subsection{\label{sec:Cauchy_inner}The Cauchy horizon and the inner horizon}

\begin{figure}
\begin{center}
\includegraphics[scale=0.85]{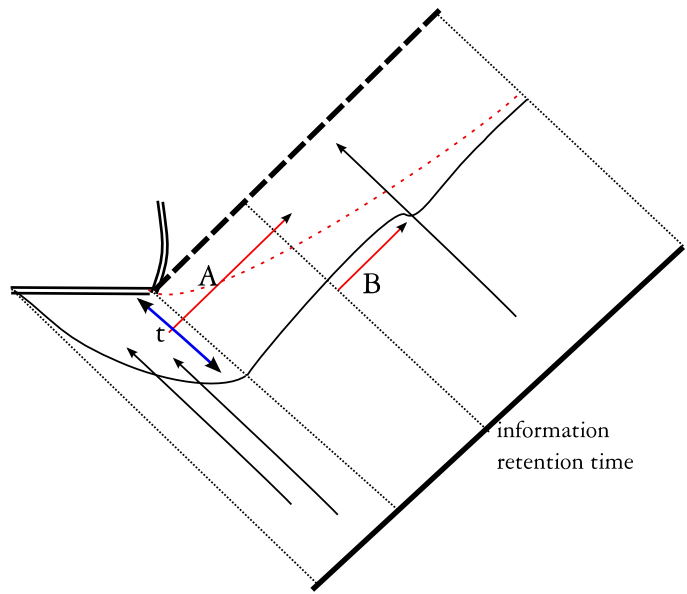}
\caption{\label{fig:complementarity_Cauchy} As long as the initial charge is small enough, we can assume that the information retention time can be located as seen in this figure. From previous simulations \cite{HHSY}, we justified the idea that some signals or observers can penetrate the inner horizon. And, it is clear that the in-falling matter has enough time to send a signal ($A$) in the outgoing direction. Thus, the duplication experiment may be possible inside of the black hole.}
\end{center}
\end{figure}

Traditionally, according to the principle of complementarity, the inner horizon was regarded as harmful to the complementarity \cite{Ge:2005bn}\cite{Thorlacius}\cite{Ge};
but the inner horizon is problematic since it is almost the same as the Cauchy horizon,
and the Cauchy horizon itself means that its beyond is not well defined.
So, in many contexts, the inner horizon is regarded as a kind of singularity \cite{Bonanno:1994ma}\cite{Ori}.
Moreover, if we ignore the Hawking radiation, this seems to be true from the mass inflation before the inner horizon \cite{HodPiran}.
However, as we discussed in the previous chapter, if we include the Hawking radiation, the whole situation becomes different.

We assume that the Hawking radiation is dominant over the pair-creation;
so, the pair creation effect will not be dominant until the information retention time.
Then, it can be justified to use Figure \ref{fig:complementarity_Cauchy}.
We can assume that the information from the Hawking radiation can be carried by $B$.
One potential problem is whether the sending of $A$ is possible or not; in a neutral case, there is no time to send the signal.
However, in our case, it seems to be possible.

Let's check more details. In Figure \ref{fig:complementarity_Cauchy}, distance $t$ should be large enough to send a signal.
If $t$ is small, from the uncertainty relation $\Delta E \Delta t \sim 1$, we need quite a large energy. If the energy is larger than the mass of the black hole itself, the duplication experiment itself is impossible \cite{inforretention}\cite{inforretention2}.
However, in Figure \ref{fig:complementarity_Cauchy}, $t$ is the order of the black hole radius; thus its inverse, or required energy, is quite a bit smaller than the mass of the black hole itself. Thus, nothing will prevent the duplication experiment.

Moreover, we know that the inner horizon is regular and stable for perturbations \cite{HHSY};
thus, penetrations of $A$ and the observer are possible.
Thus, the duplication experiment will be possible in this case.

\section{\label{sec:HM}Horowitz-Maldacena's proposal as a selection principle}

\begin{figure}
\begin{center}
\includegraphics[scale=0.6]{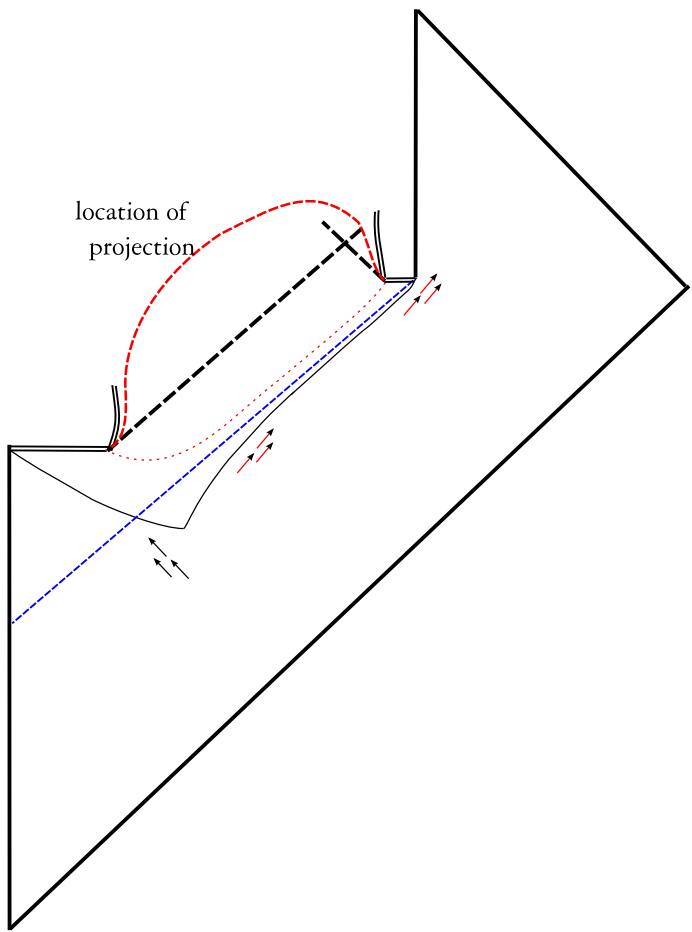}
\caption{\label{fig:charged_projection} A reasonable location of projection. It may happen around the trans-Planckian curvature region; even though we assume a large $N$, this region does exist.}
\end{center}
\end{figure}

\begin{figure}
\begin{center}
\includegraphics[scale=0.6]{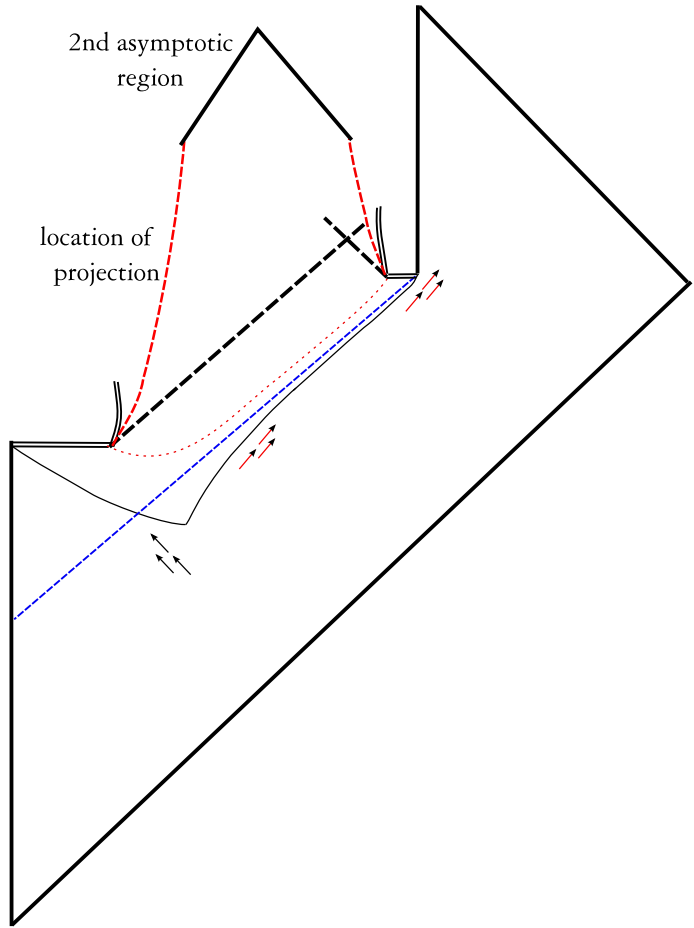}
\caption{\label{fig:charged_projection_2} If there is a second asymptotic region, the Horowitz-Maldacena proposal (and, also, the selection principle) will not work. Our simulation cannot rule out this possibility.}
\end{center}
\end{figure}

If the black hole complementarity principle is a fundamentally true idea, then how can we understand this possibility?
Some people think that because of this potential problem in complementarity \cite{Ge:2005bn}\cite{Thorlacius}, or in the context of the strong cosmic censorship \cite{cc}\cite{Poisson:1997my}, the inner horizon must be singular \cite{cc}\cite{Poisson:1997my} or there will be no inner horizon in real situations \cite{Thorlacius}\cite{FKT}.
In some models of dilaton black holes \cite{FKT}\cite{Garfinkle:1990qj}, there may be no inner horizon; but it would not be a generic phenomena \cite{Ori:2006vv}.
And, if there is an inner horizon, it can be regularized by assuming large $N$.
Thus, \textit{these objections to the inner horizon do not completely resolve the problem.}

One possibility is that we cannot observe the information from the Hawking radiation (i.e., $B$)
unless its free-falling counterpart touches the singularity.
Original inventors of the complementarity idea did not assume this; thus, this is a new assumption.
If this is true, then we cannot compare $A$ and $B$, since $A$ did not touch the singularity.
We will call it \textit{a selection principle} between the singularity and the Hawking radiation generating surface.

One may guess how to give a correlation between the singularity and the outer horizon (or, the Hawking radiation generating surface).
If we trust that the Hawking radiation is a local effect near the outer horizon,
there is no way to give a space-like correlation between the singularity and the outer horizon.

Note that the inventors of the complementarity principle tried to include some non-local effects to describe the inside and the outside of a black hole at the same time \cite{nonlocal}; because if one wants to describe them at the same time, the local quantum field theory must break down. However, this is not the selection principle between the free-falling part and the Hawking radiation; since they just assumed that, \textit{for a single observer}, there is no violation of natural laws \cite{nonlocal}. However, in the case of a charged black hole, we have argued that a single observer can observe a duplication of information; thus, \textit{this selection principle is a stronger assumption than the non-local effects.}

However, we may use quantum teleportation to realize the selection principle.
We may need to assume the collapse of the wave function,
and this may violate the unitarity of quantum mechanics.
Anyway, if we assume the collapse of the wave function near the singularity,
it may be possible to implement the selection principle in a natural, i.e., moderately unitary, way.
Horowitz and Maldacena studied this mechanism \cite{hmproposal}, and the authors will call it Horowitz-Maldacena's proposal.

According to Horowitz-Maldacena's proposal \cite{hmproposal}, we need to assume some facts:

\begin{itemize}
  \item The unique boundary condition to the singularity; also, the idea that there is no asymptotic region inside of the black hole.
  \item Outgoing Hawking radiation is maximally entangled to its ingoing counterpart.
\end{itemize}

Then, one of the pair falls into the singularity, and because of the boundary condition, it is projected there.
Then, we can easily check the unitarity of total time evolution, and it seems to act like quantum teleportaion.
Of course, those assumptions are not proven: the unique boundary condition may not work, there may be the second asymptotic region,
or the entanglement can be broken by some effects.

Although there are some problems, if this proposal is true, it will rescue the complementarity principle in a charged black hole.
Because if $A$ did not touch the singularity, $B$ cannot be related to $A$ by projection.

Notice that the location of the projection or boundary condition was not clear for a charged black hole.
Horowitz and Maldacena \cite{hmproposal} just guessed that the projection would happen at the inner Cauchy horizon.
But, according to our simulation results, the location of projection (or the singularity) and the Cauchy horizon are not same;
thus, we need to specify the location.
We observed that the inner horizon is regular,
and there is no reason to think that we must assume a kind of quantum gravitational boundary condition on the inner horizon.
Moreover, the Cauchy horizon itself seems to be regular by assuming a large $N$.
Then, the only possibility is around the central singularity.
Or, if one assumes that the Cauchy horizon is affected by some strong gravitational effect from the singularity \cite{Hawking:1992ti},
then the area around the Cauchy horizon maybe acts as the projection surface (Figure \ref{fig:charged_projection}).
This is one of the interesting results of our analysis on the causal structure of a dynamical charged black hole,
since this corrects a traditional misconception on the location of projection;
it was regarded as a null surface of the Cauchy horizon \cite{hmproposal}, but this is not clear now.

In conclusion, we can say that the Horowitz-Maldacena proposal or a kind of selection principle
seems to be more fundamental than the complementarity principle.
This conclusion is quite different from the previous opinions of other authors \cite{hmproposal2}.
The authors claim that we should require such kinds of proposals to work with the idea of complementarity.

There are some opinions that the Horowitz-Maldacena proposal cannot recover all information \cite{hmproposal2}\cite{hmproposal3}.
Moreover, the proposal assumes the inside structure of a black hole:
if there is no singularity (including a curvature singularity) inside of the black hole, complementarity will not work \cite{YZ},
and, if the causal structure contains the second asymptotic region inside of a black hole (Figure \ref{fig:charged_projection_2}),
this proposal cannot work.

\section{\label{sec:dis}Discussion}

We have explored the causal structure of a dynamical charged black hole.
We observed that the inner horizon that is locally defined can be different from the Cauchy horizon, which is globally defined.
From the confirmation of numerical simulations, we could conclude that the complementarity principle can be violated from a duplication experiment.
However, if we assume a selection principle between the singularity and the Hawking radiation generating surface,
the apparent problem may be resolved.
And, the authors argued that the Horowitz-Maldacena proposal can work as such a selection principle.

Note that, our argument used many parts from \cite{HHSY}, but the basic idea does not need the details of the paper.
First, in fact, we need only the causal diagram that connects from the mass inflation scenario to the near extreme one
(Figure \ref{fig:charged_with_mass_inflation2}); and, we need not the whole diagram for the duplication experiment, i.e., Figure \ref{fig:charged}.
Then, the only possible way is that the inner horizon bends space-like direction;
thus, naturally, there is difference between the inner horizon and the Cauchy horizon.
This was confirmed by not only the authors, but also another author, e.g., \cite{SorkinPiran2}.
Second, we need to regularize the curvature problem around the inner horizon;
as we commented, we can easily see that the curvature is not infinite at the inner horizon,
and the large curvature can be regularized by large $N$. Those ideas were sufficiently explained in this paper.

Our conclusion invokes interesting questions.
The membrane paradigm or the holographic principle is in debt to observer complementarity,
which is a generalized version of black hole complementarity.
If one thinks the membrane paradigm or the holographic principle, he or she will choose the outside observer.
Of course, we cannot forget on the free-falling observer, and only if we believe observer complementarity, it is justified to forget the free-falling observer.

However, for a charged black hole, complementarity would require an assumption on the inside structure of the black hole
and/or a speculative selection principle like the Horowitz-Maldacena proposal.
Thus, although we can see the charged black hole in a way that complementarity is working apparently,
we have to do more study to justify additional assumptions in the fundamental level.
If the Horowitz-Maldacena proposal is incomplete, complementarity may not work.
Unfortunately, there are some potential problems with Horowitz-Maldacena's proposal:
suppositions (e.g., collapsing of the wave function) of the proposal may not be true, the proposal may imply information loss (e.g., if there are strong interactions between the in-falling matter and the in-going Hawking radiation), and if there is no singularity (e.g., for a regular black hole), the proposal cannot work.
Moreover, if beyond the Cauchy horizon has a second asymptotic region, it will not work.

(Note that, there is a recent work of the authors on related issues. See, \cite{YZ2}.)

Originally, the principle of black hole complementarity was introduced to resolve the information loss paradox.
According to our analysis, we conclude that this can give a partial resolution to the problem
but has the burden of requiring supplement from another (maybe incomplete) idea like Horowitz-Maldacena's proposal.
Thus, in some sense, other ideas, which would be in debt to complementarity, may have significant problems.
Thus, dynamical black holes including regular black holes must be considered more carefully.

\acknowledgments{
The authors would like to thank Ewan Stewart for discussion and encouragement.
This work was supported by BK21 and the Korea Research Foundation Grant funded by the Korean government (MOEHRD; KRF-2005-210-C000006, KRF-2007-C00164).
The authors would also like to thank the Korea Science and Engineering Foundation(KOSEF) for a grant funded by the Korean government (No. R01-2005-000-10404-0).
}

\appendix

\section{\label{appa}A brief introduction to basic setup of HHSY}

We will briefly introduce the basic schemes of \cite{HHSY} that were used several times in this paper.

We assumed a complex massless scalar field $\phi$ that is coupled from the electromagnetic field $A_{\mu}$ \cite{Hawking:1973uf}:
\begin{eqnarray} \label{Lagrangian}
\mathcal{L} = - (\phi_{;a}+ieA_{a}\phi)g^{ab}(\overline{\phi}_{;b}-ieA_{b}\overline{\phi})-\frac{1}{8\pi}F_{ab}F^{ab},
\end{eqnarray}
where $F_{ab}=A_{b;a}-A_{a;b}$, and $e$ is the unit charge.
For convenience, spherical symmetry is a useful assumption:
\begin{eqnarray} \label{double_null}
ds^{2} = -\alpha^{2}(u,v) du dv + r^{2}(u,v) d\Omega^{2},
\end{eqnarray}
where we used the double--null coordinate (our convention is $[u,v,\theta,\phi]$) \cite{Hamade:1995ce}.
Moreover, since the electromagnetic field is a gauge field, we can fix a gauge $A_{\mu}=(a,0,0,0)$ \cite{OrenPiran}.
Then, those assumptions give the Einstein tensor $G_{\mu\nu}$ and the stress-energy tensor $T_{\mu\nu}$ components,
as well as equations of motion for the scalar field and the electromagnetic field.

Moreover, to include the Hawking radiation, we need to introduce the renormalized stress-energy tensor $\langle T_{\mu\nu} \rangle$. We used 2-dimensional results \cite{Davies:1976ei}, which are divided by $4 \pi r^{2}$; this is a reasonable assumption for the spherically symmetric case \cite{SorkinPiran2}:
\begin{eqnarray} \label{semiclassical}
&& \langle T_{uu} \rangle = \frac{P}{4\pi r^{2} \alpha^{2}}(\alpha \alpha_{uu} - 2 {\alpha_{u}}^{2}),
\nonumber \\
&& \langle T_{uv} \rangle = \langle T_{vu} \rangle = -\frac{P}{4\pi r^{2} \alpha^{2}}(\alpha\alpha_{uv}-\alpha_{u}\alpha_{v}),
\nonumber \\
&& \langle T_{vv} \rangle = \frac{P}{4\pi r^{2} \alpha^{2}}(\alpha \alpha_{vv} - 2 {\alpha_{v}}^{2}),
\end{eqnarray}
($B_{a}$ is a partial derivative of a function $B$ with respect to $a$) where $P \propto Nl_{pl}^2$; $N$ is the number of massless scalar fields generating the Hawking radiation, and $l_{pl}$ is the Planck length.
By changing $P$, we can tune the strength of quantum effects; although we fix $P$, by tuning $N$, we can tune the Planckian cutoff \cite{HHSY}.

Typically, the mass function increases exponentially around the inner horizon; its order is $\exp(\kappa v) \sim \exp(M^{2})$, since typical surface gravity of the inner horizon is on the order of $1/M$, and the typical time scale is its life time $v \sim M^{3}$. In simulations, literal values of curvatures were on the order of $\exp(100)$, where typical radius was on the order of $10$ \cite{HHSY}. By re-scaling of the number of massless fields $N$ or Planck length $l_{pl}$, its physical length should be changed, and the curvature should be re-scaled by $R/N$. Thus, a large number of $N$ will resolve the trans-Planckian problem.

Finally, we will use the Einstein equation:
\begin{eqnarray} \label{Einstein}
G_{\mu\nu}=8\pi (T_{\mu\nu}+\langle T_{\mu\nu} \rangle).
\end{eqnarray}
Then, we have all the equations we need to solve. For complete equations, see \cite{HHSY}.

For initial conditions, we push the charged field configuration at the initial surface:
\begin{eqnarray} \label{s_initial}
\phi(u_{i},v)=\frac{A}{\sqrt{4\pi}} \sin ^{2} \left( \pi \frac{v-v_{i}}{v_{f}-v_{i}} \right) \exp \left( \pm 2 \pi i \frac{v-v_{i}}{v_{f}-v_{i}} \right),
\end{eqnarray}
where $u_{i}$ and $v_{i}$ represent the initial retarded and advanced time, and $v_{f}$ indicates the end of the pulse in the initial surface.
Finally, after fixing some parameters, we will leave the undetermined three parameters: the unit charge $e$, the strength of Hawking radiation $P$, and the amplitude of the field $A$. And, by choosing three parameters, we determine one specific simulation.

\end{document}